# The advanced setup for synthesis of composite long-length superconducting tape of $Nb_3Sn$

B. G. Lazarev, B. V. Borts, P. A. Kutsenko, O. P. Ledenyov, E. Yu. Roskoshnaya, V. I. Sokolenko

*National Scientific Centre Kharkov Institute of Physics and Technology, Academicheskaya 1, Kharkov 61108, Ukraine.*

The description of design of advanced technological setup for the synthesis of composite long-length superconducting tape on the basis of $Nb_3Sn$ is presented. The design of setup has allowed us to produce the $Nb_3Sn$ tape with the record magnitudes of density of critical current $J_c$ in the magnetic fields $H$ of above $10\ T$ ($H > 10\ T$). The high current density conductivity property is reached as a result of both the optimization of technological process of the $Nb_3Sn$ tape synthesis and the increase of thickness of superconductor layer, which is characterized by the homogeneous physical properties.

## Introduction

The superconducting composites with the crystal structure *A15* are characterized by the high magnitudes of critical parameters of superconductor (the critical temperature $T_c$, second critical magnetic field $H_{c2}$, density of critical current $J_c$) [1], allowing their application to develop the laboratorial superconducting solenoids and complex superconducting magnetic systems for the use in the high magnetic fields for various purposes [2]. In the class of the superconductors *A15*, the composite superconductor of $Nb_3Sn$ is well researched. Presently, a big number of the liquid-phase and solid-phase technologies for the synthesis of the superconducting tapes and multifilament wires on the base of superconducting composite of $Nb_3Sn$ are developed (see, for example, [1, 3, 4]). At *NSC KIPT*, the composite long-length superconducting tape of $Nb_3Sn$ with the record magnitude of critical current density $J_c \sim 3 \cdot 10^{-2} \cdot J_0$, where $J_0 = 1,8 \cdot 10^8\ A \cdot cm^{-2}$ is the limiting magnitude of density of critical current, corresponding to the reach of critical velocity by the electrons in superconducting condensate [5], was synthesized. In agreement with the research results in [5], the creation of the current caring microstructure in the superconductor of $Nb_3Sn$ is a spontaneous process, connected with the self organization of this structure.

The high magnitudes of critical current density $Jc$ can be reached in the composite long-length superconducting tape on the basis of $Nb_3Sn$, in which the density of critical current doesn't depend on the thickness of the superconducting tape layer at the high magnitudes of current. The superconducting tape was fabricated with the use of specially designed and developed technological setup, which utilizes the liquid–phase method of phase formation of $Nb_3Sn$.

The layer of $Nb_3Sn$ with the improved superconducting characteristics was placed on the thin ribbon of the pure *Niobium* or *Niobium*, which was alloyed with ~1,5% *Zirconium*, using the thermo-diffusion process. During this process, the original ribbon of *Nb* interacts with the melted *Tin* or *Tin-Copper* alloy, undergoing the thermal treatment and oxidation process. The optimal temperature for the superconducting layer synthesis was found, using the diagram of phase state of *Niobium-Tin* system and going from the precisely measured dependence of the critical current density on the temperature.

The development of setup resulted in the solution of the problem of optimization of synthesis process toward the fabrication of better quality superconducting tape with the thermo-diffusion layer of $Nb_3Sn$ with the high critical current density $J_c$ characteristics in high magnetic fields $H$ of $10\ T$ and above. The main challenge was to minimize the origination of accompanying non-superconducting intermetallic phases $Nb_6Sn_5$ and $NbSn_2$ in the *Niobium-Tin* system by finding an appropriate operational temperature range for the plant. These phases of $Nb_6Sn_5$ and $NbSn_2$ synthesized at the temperatures of below $906°C$ and $860°C$ correspondingly. It was understood that, at long enough exposures of plated tape to these temperatures, the described inter-metals of $Nb_6Sn_5$ and $NbSn_2$ synthesize in the form of thin interlayers between the $Nb_3Sn$ crystal grains, resulting in the limitation of critical current density. The distribution of temperatures inside the oven was appropriately adjusted with the aim to avoid the $Nb_6Sn_5$ and $NbSn_2$ non-superconducting phases origination in the composite long-length superconducting tape on the basis of $Nb_3Sn$.

In this research paper, the detailed description of design of the improved advanced experimental setup for the synthesis of the composite long-length superconducting tape of $Nb_3Sn$ is presented.

# Principal scheme of advanced technological set up for Nb₃Sn tape synthesis

The thin ribbon of pure *Niobium* with the thickness of 12…14 μm or the thin ribbon of *Niobium*, which was alloyed with the *Zirconium* (1,6...1,7 %) with the thickness of 17…20 μm represented the main source material for the synthesis of composite long-length superconducting tape on the basis of $Nb_3Sn$. The fabrication of the thin ribbon of pure *Niobium*, included a number of functional operations at special setups, devised for the cutting, ultrasonic purification, and ablution. After the preparation stages, the deposition of *Tin* layer or *Tin-Copper* alloy layer was conducted at the technological setup, which is shown in Figs. 1, 2.

Fig. 1, the setup, which includes the vacuum chamber with the related equipment and pumping system, is shown. It automatically operates and uses of ribbons with the width of *10 mm*. The pumping system consists of the mechanical vacuum pump *ВН-2МГ* with the limiting vacuum of $3 \cdot 10^{-3}$ *Torr* (16), providing the preliminary pumping of gas up to the first vacuum, and also the pumping by the diffusion oil pump *ММ-40* (19) and by the absorption pump (20), refrigerated in the liquid *Hydrogen*, which allows to pump the gas admixtures, including the *Hydrogen*, exhaling during the process of *Tin* plating of *Niobium* or *Niobium-Zirconium* ribbons.

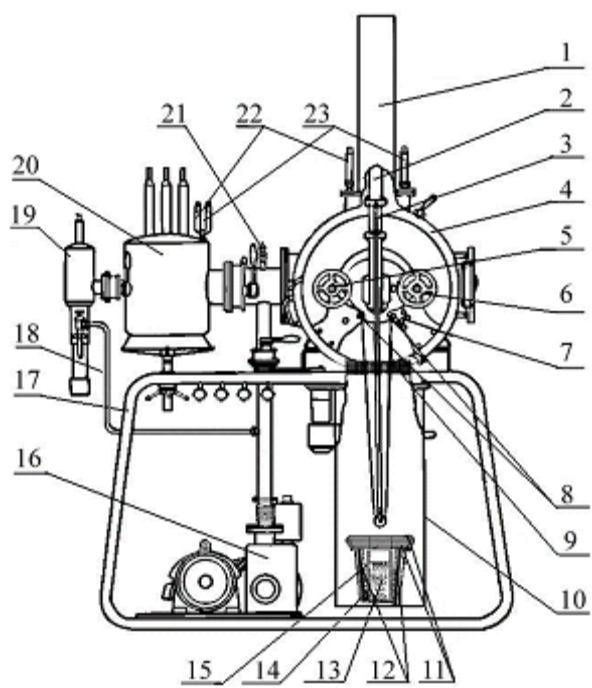

*Fig. 1. Scheme of setup to synthesis composite long-length superconducting tape on basis of Nb₃Sn (side appearance): 1 – the cover of top chamber; 2 – the direction guider for descending and lifting of quartz shoot with roll; 3 – the quartz shoot with quartz horn; 4 – the top part of operating chamber; 5 – loading bobbin with pure Niobium ribbon; 6 – the receiving bobbin for Tin plated Nb₃Sn tape; 7 – the system of rolls for tape rolling; 8 – the direction guiding rolls; 9 – the system of screens for optimization of thermal losses; 10 – the bottom part of operating chamber; 11 – the heat shielding (top and side) screens; 12 – the Molybdenum heater in oven; 13 – the melted Tin; 14 – the quartz skirt, in which melted Tin is situated; 15 – the oven to melt Tin; 16 – the mechanic vacuum pump ВН-2МГ; 17 – the pipe to refrigerate setup; 18 – the vacuum pipeline for diffusion pump ММ-40 (19) pumping by mechanic vacuum pump ВН-2МГ (16); 19 – the diffusion oil pump ММ-40 with vacuum trap; 20 – the high vacuum absorption pump; 21 – the vacuum tap; 22 – the manometer bulbs ЛТ-2; 23 – the manometer bulbs ЛМ-2.*

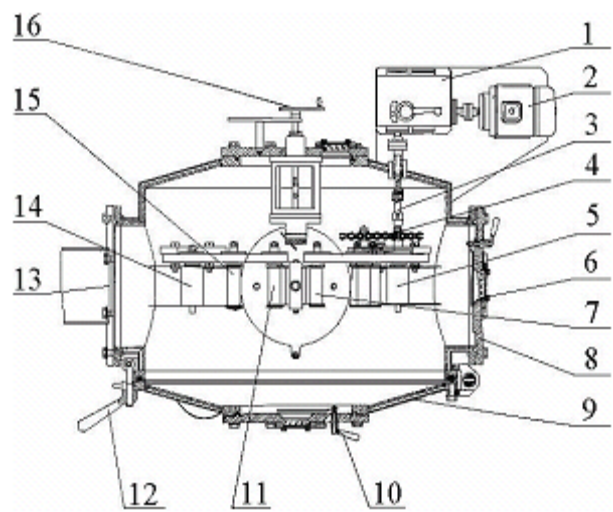

*Fig. 2. Scheme of setup (atop appearance): 1 – the speed reducer; 2 – the electrical motor АОЛ-012/4; 3 – the speed reducer's axle; 4 – the circuit, activating the drawing mechanism; 5 – the bobbin to accumulate plated tape; 6 – the sighting window; 7 – the directing roll; 8 – the side flange; 9 – the shell of upper part of operating chamber; 10 – the handle of shielding screen of sighting window; 11 – the directing roll; 12 – the bolt to close the opening cover in the upper part of operating chamber; 13 – the flange to connect with vacuum system; 14 – the loading bobbin with Niobium ribbon; 15 – the directing roll; 16 – the mechanism of shoot down activation.*

The working chamber consists of the upper (4) and lower (10) parts. It is completed in the form of a cylinder made of the stainless steel. On the left side, the flange of working chamber serves to make a connection to the pumping system. The vacuum tap (21), which regulates the inflow of air to the chamber, is situated on the same flange.

The oven (15) to melt the *Tin* or *Tin-Copper* alloy is placed in the center of the lower chamber (10). This oven (15) has a system of screens (11) in its upper part, decreasing the thermal radiation by the oven. The heater (12) in the oven is made of *Molybdenum* wire with the diameter 1,5 *mm*, which is reeled on the pipe with the cone shape with the decreasing diameter toward the oven's bottom.

The temperature measurements of the oven and melted *Tin* are performed by the two *Chrome* thermo-pairs. The first thermo-pair is situated between the quartz skirt (14) and the oven (15) in a special quartz tube with the curved form, which prevents the deposition of platting allow on the thermo-pair. The second thermo-pair is situated in the quartz tube of shoot (3), allowing the temperature measurement in close proximity to the place of contact between the pure *Niobium* ribbon and the quartz roll in the melted alloy.

The chamber (10), in which the oven is situated, can be moved down with the help of mechanical jack, and then be moved aside to load the quartz skirt with the alloy, during the retracting operation and other technological operations.

The top part of chamber (4) is covered by the folding cover with sighting window, which is shielded by a screen. The cover can be closed with the help of three adding bolts. There are the loading bobbin with cleaned ribbon (5), receiving bobbin with plated tape (6), drawing rolls (7) and directing rolls (8) in this part of operational chamber.

The top part of operational chamber (see Fig. 2) (top appearance) has the three sighting windows, covered by the screens, making it possible to conduct an observation during the process of tape movement.

The plated ribbon is putted to the melted *Tin* or *Tin-Copper* alloy by the shoot (3) (see Fig. 1) with the quartz roll at the end. The upper end of shoot is fixed on the mechanism, sliding toward the direction pointer, situated in the top part of the chamber. The movement of quartz shoot is provided with the help of mechanism with the jagged transmission, having the circular scale, allowing to precisely douse the quartz tube with plated tape down to the needed deepness (see Fig. 2, (16)).

The device to roll and drag the ribbon represents the system described in Fig. 2. At the left part of chamber, there is a loading bobbin with the original pure *Niobium* ribbon, which is placed on the pillow-block with the aim to decrease the friction. The degree of bobbin's clench is regulated by the spring, fixed by the two nuts. There is directing roll near to the loading bobbin. This roll rotates on the pillow-block. The ribbon, passing through the roll, upper protective screens, screens of oven, quartz roll submerged in the plating melting pot, comes to the roll (7) of intake part of drawing mechanism. After this, the plated ribbon passes through the system, consisting of four rolls, rotating with the constant velocity and providing the engagement within it. These rolls are activated with the help of speed reducer (3) with the shaft introduced to the chamber through the vacuum packing. The speed-reducer is rotated by the electromotor *АОЛ-012/4*. There is a switch on the speed-reducer, allowing to draw the tape with the velocity 5,2; 10 and 15 *m per hour*.

The ribbon, covered by the alloy, passes the system of four rolls, coming to the receiving bobbin, which is rotated by the chain transmission, which connects this bobbin with the system of drawing rolls. The rotation of receiving bobbin is realized, using the «slipping regime» of operation with the purpose to decrease the stretch of loaded tape. The vacuumed electrodes are used for the power supply to the heater in the oven and the outputs of thermo-pairs from the bottom and top parts of operation chamber. The vacuum packing of all the inputs is made with the help of rubber strips, and the electric isolation is achieved due to the application of the *Teflon* sockets. The operational chamber has the double walls with the refrigerated water pumping to protect the vacuum packing from a possible overheating.

The power feed with the voltage of *220 V* comes to the stabilizer from the distribution dash, and then to the transformer *АОСК-10/09*, which is connected with the electrical oven in the setup.

The setup is characterized by the following parameters. The volumes of the upper and lower chambers are *154* and *125 $dm^3$* correspondingly. The full preparation time from the switching on the vacuum pumps to the beginning of actual operation is *1,5 hours*. The power of oven is *3,5 κW* at the temperature *1000 °C*. The receiving bobbin can accommodate up to *400 m* of the *Tin* plated composite long-length superconducting tape on the basis of $Nb_3Sn$ with the width of *10 mm* and thickness of *40 μm*. The removable bobbins for the tapes with the width of *40* and *80 mm* are present in the setup.

The subsequent thermal treatment of the *Tin* plated composite long-length superconducting tape on the basis of $Nb_3Sn$ was done at a separate setup with the special configuration of temperature field, aiming to minimize the time duration of the $Nb_3Sn$ tape exposure to the high temperatures at which the inter-metals of $Nb_6Sn_5$ and $NbSn_2$, which have a negative influence on the transport current properties, can be originated. These and some other improvements allowed to obtain the homogenous thick layer of $Nb_3Sn$ on the surface of *Nb* ribbon, resulting in the synthesis of composite long-length superconducting tape on the basis of $Nb_3Sn$ with record magnitudes of the critical current density.

In Fig. 3, the measured dependences of the magnitude of critical current in the $Nb_3Sn$ tape at *T=4,2 K* in the magnetic field *6 Tesla* on the temperature of the *Tin* melt, measured after its plating, and also after the thermal treatment at *T = 905 °C* during *20 hours*., are shown.

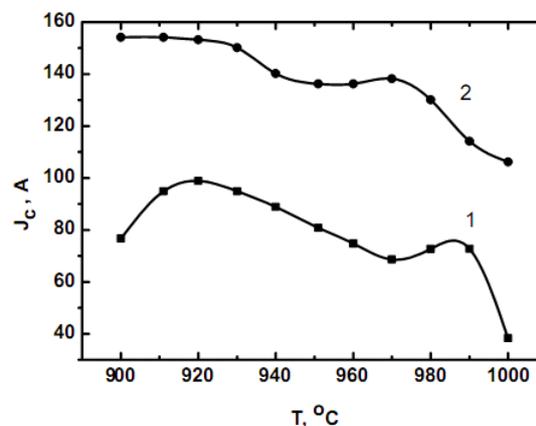

*Fig. 3. Dependences of magnitudes of critical current in superconducting $Nb_3Sn$ tape (related to 1 cm of its width) on temperature of Tin melt: 1 – initial Tin plated Nb ribbon with the layer of $Nb_3Sn$; 2 – thermally treated Tin plated Nb ribbon with the layer of $Nb_3Sn$.*

It can be seen that that the optimal temperature regime of plating process, which gives an opportunity to synthesize the $Nb_3Sn$ composite long-length superconducting tape with best physical characteristics, is in the range of temperatures between *900…920 °C*.

After the completion of optimization of technological process, including the *Zirconium* oxidation, it was found that the use of the *Niobium-Zirconium* ribbon, plated by the *Tin-Copper* alloy, results in a significant increase of the magnitude of critical current density up to the record value $Jc = 10^6$ *A·cm$^{-2}$* in the $Nb_3Sn$ composite long-length superconducting tape in the magnetic field *H* of *10 Tesla*.

## Conclusion

The design of advanced technological setup for the $Nb_3Sn$ tape synthesis allowed us to fabricate the composite long-length superconducting tape on the basis of $Nb_3Sn$ with the record magnitudes of the critical current density $J_c = 10^6$ *A·cm$^{-2}$* in the magnetic fields above *10 Tesla* at the *Helium* temperatures. The high transport current property in $Nb_3Sn$ tape was reached due to an optimization of technological process, including the deposition of *Tin* or *Tin-Copper* melting on the surface of original $Nb_3Sn$ tape and the application of thermo-diffusion synthesis of the $Nb_3Sn$ superconductor layers with the thickness up to *6 μm* at every side of tape, characterized by the high homogeneous physical properties. The produced $Nb_3Sn$ tape was used for the design of series of superconducting solenoids with the high magnetic fields (*H > 10 Tesla*).

This research is completed in the frames of the fundamental and applied superconductivity research program at the National Scientific Centre Kharkov Institute of Physics and Technology (*NSC KIPT*) in Kharkov in Ukraine.

Authors sincere thank Lyubov S. Lazareva, V. A. Poltavets, A. L. Donde and all the engineers at the *Schubnikov* cryogenic laboratory under the supervision by A. P. Sheinin, who took an active part in the research work towards the design and development of advanced technological setup for the $Nb_3Sn$ tape synthesis.

This article was published in the *Problems of Atomic Science and Technology* (*VANT*) in 2009 [6].

*E-mail:   ledenyov@kipt.kharkov.ua

———————

1. V. M. Pan, V. G. Prohorov, A. S. Shpigel, Metalophysics of Superconductors, *Naukova Dumka*, Kiev, Ukraine, 189 p., 1984.
2. M. Wilson, Superconducting Magnets, *Mir*, Moscow, Russian Federation, 407 p., 1985.
3. B. G. Lazarev, V. M. Pan, On the Perspectives of Increase of Critical Parameters of Superconductors, *Metalophysics*, vol. **1**, issue 1, pp. 52-62, 1979.
4. V. P. Korzhov, Methods of Synthesis of Superconducting Materials on the Base of Intermetallic Compounds with the A−15 Structure: Review, *Problems of Technical Superconductivity*, Chernogolovka, Russian Federation, pp. 5−43, 1984.
5. B. G. Lazarev, P. A. Kutsenko, L. S. Lazareva, B. K. Pryadkin, N. A. Chernyak, On the Nature of Limiting Critical Current Density in Nb$_3$Sn Layers, *Metalophysics*, vol. **12**, no. 3, pp.18-24, 1990.
6. B. G. Lazarev, B. V. Borts, P. A. Kutsenko, O. P. Ledenyov, E. Yu. Roskoshnaya, V. I. Sokolenko, *Problems of Atomic Science and Technology* (*VANT*), no. 6(18), pp. 111-114, ISSN 1562-6016, 2009 (in Russian).
http://vant.kipt.kharkov.ua/ARTICLE/VANT_2009_6/article_2009_6_111.pdf .